\documentclass[aps,prl,superscriptaddress,twocolumn]{revtex4}

\usepackage{graphicx}
\usepackage{amssymb}
\usepackage{amsmath}
\usepackage{amsfonts}
\usepackage{amssymb}
\usepackage{graphicx}
\usepackage{hyperref}
\usepackage{color}

\begin{document}

\title{Dipole oscillations of confined lattice bosons in one dimension}

\author{Simone Montangero}
\affiliation{NEST-CNR-INFM and Scuola Normale Superiore, I-56156 Pisa,
  Italy}
\affiliation{Institut f\"ur Quanteninformationsverarbeitung, Universit\"at Ulm, 
D-89069 Ulm, Germany}
\author{Rosario Fazio}
\affiliation{NEST-CNR-INFM and Scuola Normale Superiore, I-56156 Pisa, Italy}
\affiliation{International School for Advanced Studies (SISSA), I-34014 Trieste, Italy}
\author{Peter Zoller}
\affiliation{Institute for Theoretical Physics, University of
Innsbruck,  A-6020 Innsbruck, Austria}
\affiliation{Institute for Quantum Optics and Quantum Information,
A-6020 Innsbruck, Austria}
\author{Guido Pupillo}
\affiliation{Institute for Theoretical Physics, University of
Innsbruck,  A-6020 Innsbruck, Austria}
\affiliation{Institute for Quantum Optics and Quantum Information,
A-6020 Innsbruck, Austria}


\date{\today}

\begin{abstract}
We study the dynamics of a non-integrable system comprising interacting cold bosons trapped
in an optical lattice in one-dimension by means of exact time-dependent numerical DMRG
techniques. Particles are  confined by a parabolic potential, and dipole oscillations
are induced by displacing the trap center of a few lattice sites. Depending on the system
parameters this motion can vary from undamped to overdamped. We study the dipole oscillations
as a function of the lattice displacement, the particle density and the strength of
interparticle interactions.
These results explain the recent experiment C.~D.~Fertig  {\em et al.}, Phys. Rev. Lett. 94, 120403 (2005).
\end{abstract}

\maketitle

Recent experiments with cold atoms~\cite{pezze04,kinoshita06,stoeferle04,
fertig05} have provided realizations of non-equilibrium quantum many-body
systems, allowing to address a number of fundamental questions. For example,
the integrability of a many-body system has been demonstrated in
Ref.~\cite{kinoshita06}, via the inhibition of thermalization in  a
one-dimensional Bose gas, which opened the way to theoretical studies of the
relaxation dynamics of non-equilibrium many-body systems~\cite{kollath07}. The
dynamics of non-integrable systems  
has been recently explored experimentally in Refs.~\cite{stoeferle04,fertig05}
using interacting cold bosonic atoms trapped in an array of one-dimensional
optical lattices and confined by a parabolic potential. Dipole oscillations
were induced by displacing the center of the parabolic potential, and the
dipole dynamics was studied by monitoring the position of the center of
mass. A sudden transition from a regime of undamped motion to a regime of
strongly damped motion was observed on increasing the lattice depth. 
Since damping of the center of mass oscillations is due to excitations in the
optical lattice, the results obtained in~\cite{stoeferle04,fertig05} have
provided precious diagnostic of the dynamical correlations of the many-body
system, and thus have stimulated considerable 
theoretical interest~\cite{Polkovnikov,RuostekoskiPRL05,PupilloNJP06}. 

Good agreement with the experimental results in~\cite{fertig05} has been obtained in the regimes of very
weak~\cite{RuostekoskiPRL05} and very strong interactions~\cite{PupilloNJP06}, where mean-field
and extended fermionization techniques apply. However, it remains a
fundamental challenge to understand the dipole dynamics in the regime of
intermediate interactions, where the sudden localization transition occurs and
the subtleties of one-dimensional (1D) 
correlations do not allow (semi-)analytical treatments.  
With the aim to provide a comprehensive explanation of the experiment of Fertig {\em et al.}~\cite{fertig05}, in this letter we study
 the dipole oscillations by means of a numerically exact time-dependent
density-matrix-renormalization-group technique (tDMRG), see
also~\cite{Clark08}. We find very good agreement with the experimental results
in the interesting regime of intermediate interactions. These results
demonstrate that time-dependent numerical simulations with tDMRG have reached
the same accuracy of current experiments with cold gases in the strongly
correlated regime and thus represent a unique theoretical tool for
quantitative comparisons and predictions for experiments in the cold atoms
context. 

The experiment in~\cite{fertig05} was performed in a parameter regime where
the use of the following Bose-Hubbard Hamiltonian is microscopically
justified~\cite{Jaksch98} 
\begin{eqnarray}
H &=& -J \sum_{j}
 (b_{j}^{\dagger}b_{j+1}+\mbox{h.c.}) +\Omega \sum_j [j+ \delta(t)]^2 n_j\nonumber\\
&+&\frac{U}{2} \sum_j n_j(n_j-1).
\label{HamBHInt}
\end{eqnarray}
The first term on the r.h.s. of Eq.(\ref{HamBHInt}) describes the tunneling of bosons between
neighboring sites with rate $J$ ($j$ labels the sites on the lattice). The second term is the parabolic
potential with curvature $\Omega$; $\delta(t)$ is a sudden displacement of the trap center,
$\delta_0(t) = \delta \;\Theta(t)$ (with $\Theta(t)$ the Heaviside function), and
$n_{j} = b^{\dag}_{j} b_{j}$ is the density operator with bosonic creation (annihilation)
operators $b^{\dag}_{j}$ ($b_{i}$). The last term is the onsite contact interaction with energy
$U$~\cite{Jaksch98}, (we set $\hbar=1$).
\begin{figure}[t]
\begin{center}
\includegraphics[width=\columnwidth]{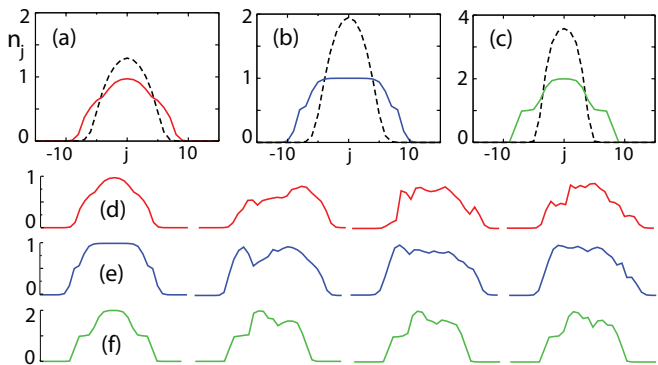}
\end{center}
\caption{Relevant density distributions [panels (a-c)], see text, and
  snapshots of the corresponding dipole dynamics [panels (d-f)].
  (a-b) Density distribution for $N=11$ and 15 particles, respectively, 
  for $\Omega/J=0.05623$. In each panel, the dashed and solid lines are
  $U/J=1$ and 20, respectively. The solid line in panel (b) corresponds to a
  Mott insulator.
  (c) Density distribution for $N=23$,
  $\Omega/J=0.4$, and $U/J=1$ and 20 (dashed and solid lines,
  respectively). The solid line corresponds to a cake-like structure. 
  (d-f) Snapshots of the density distribution for the cases
  (a-c), with $U/J=20$ and $\delta=4$, at times $tJ=0, 30, 40$ and
  50. The dynamics of a few atoms in the Mott and cake-like configurations is
  frozen, however, residual oscillations can persist  in the latter, see
  text.}
\label{fig:fig0}
\end{figure}

\begin{figure*}[htb]
\begin{center}
\includegraphics[width= 0.8 \textwidth]{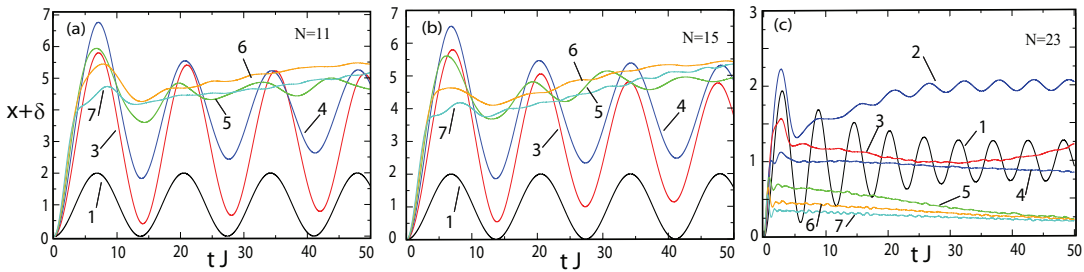}
\end{center}
\caption{Center of mass position as a function of time for the cases of
  Fig.~\ref{fig:fig0}(a-c) and $U/J=1$. The displacements $\delta$ are
  indicated in the figure. The critical displacement $\delta_{\rm c}$ equals
  $\delta_{\rm c} =6$ and 2 in panels (a-b) and (c),
  respectively.}
\label{fig:fig1}
\end{figure*}

\begin{figure*}[t]
\begin{center}
\includegraphics[width= 0.8\textwidth]{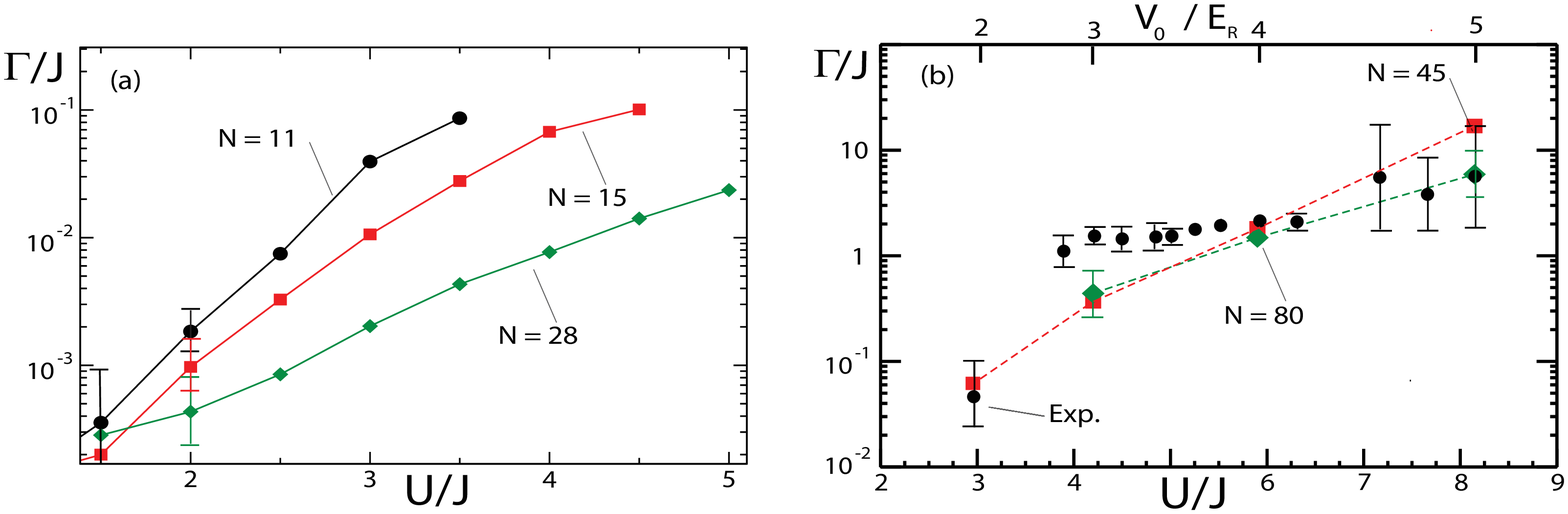}
\end{center}
\caption{(a) Numerical results for the damping rate $\Gamma$ of the dipole
  oscillations vs $U/J$ for a fixed displacement $\delta=1<\delta_{\rm c}$,
  with $\delta_{\rm c}=6$
  ($\Omega/J=0.05623$) and $N=11,15$ and 28 [cases (a-c) in the text]; (b) Damping rate
  $\Gamma$ for the experiment of Ref.~\cite{fertig05} vs $U/J$ and the lattice
  depth $V_0/E_R$. The experimental data, and the numerical results for
  $N=45,80$ are the black dots, the red squares and the green diamonds,
  respectively.}
\label{fig:fig2}
\end{figure*}

The sudden displacement on the trap center causes dipole oscillations of the bosons which
can be analyzed experimentally by monitoring the time evolution of the Center
Of Mass (COM)  $x_{\rm com}= \sum_j j \langle n_j \rangle/N$, with $N$ the number of particles.
The experiment of Ref.~\cite{fertig05} was performed on a array of one-dimensional
optical lattices 
where the number of particles in each 1D lattice varied from $  N\simeq 80$ to zero.
Thus, in order to provide a comprehensive and quantitative comparison with the
experimental data, here we analyze the dipole dynamics as a function of
$\delta$,  $U/J$, and the number of bosons $N$. 
We find that overdamped motion can occur  as a function of $\delta$ for arbitrarily
small interactions, Fig.~\ref{fig:fig1}, while in general sizeable interactions tend to extend the parameter
region where localization occurs~\cite{SmerziPRL}. For a given $\Omega/J$
 damping is found to depend exponentially on $U/J$, and to be favored for small $N$.
Figure~\ref{fig:fig2}(a), where the damping rate is shown as a function of the interaction and the number of
bosons and, most important, Fig.~\ref{fig:fig2}(b), where we compare our numerics with the
experimental data finding very good agreement in the intermediate range of
interactions, allow for a new explanation of the experiment of
Ref.~\cite{fertig05}, based on the role of lattices with different $N$. 

Three regimes are of interest for the dipole dynamics [see Fig.~\ref{fig:fig0}]: $a$) for $4J
\gtrsim \Omega(N/2)^2$ the density distribution is Gaussian or
Thomas-Fermi-like for $4J \gg U$ and $4J \simeq U$, respectively, and for
$U\gg 4J$ onsite densities are smaller than one; $b$) for $U > \Omega(N/2)^2> 4J$ a Mott
insulator with one particle per site is formed at the trap center; $c$) for
$\Omega(N/2)^2> U > 4J$ a shell structure is formed with a  density
$1<n_j\leq 2$ at the trap center, surrounded by a Mott-insulator with one particle
per site. All the situations above occur in the experiment, since $N$ varies
from one lattice to another. Therefore, in the following we are first
interested on the dynamics of model systems as those in Fig.~\ref{fig:fig0},
which exemplify all 
three cases $a$), $b$) and $c$) above while still allowing for an extensive
analysis in terms of all parameters $N, \Omega/J$ and $U/J$, and then we
address the experiment of 
Ref.~\cite{fertig05} in the most interesting regime $U/J \gtrsim 4$.

The results presented below have been obtained by means of a tDMRG algorithm with a
second order Trotter expansion of $H$, and time-steps $0.01
J$~\cite{daley04}. We take advantage of the conserved total number of
particles $N$ projecting on the corresponding subspace;
the truncated Hilbert space dimension is up to $m=100$,
while the allowed number of particles per site is $D=5$.
All results below are found to be independent of this choice.

We first focus on the dipole dynamics as a function of the trap displacement
$\delta$, in the regime of weak interactions. In this regime, mean-field
theory predicts a sudden transition between undamped and overdamped motion via a
dynamical instability at a critical displacement $\delta_{\rm c} \simeq \sqrt{2 J/
\Omega}$~\cite{SmerziPRL}. This value for $\delta_{\rm c}$ can be understood by
employing the exact solution of Eq.~\eqref{HamBHInt} in the non-interacting
limit~\cite{AnaPRA05}. For energies $E \lesssim 4J$ the single-particle eigenstates
of $H(t=0)$ are harmonic-oscillator-like modes extended around the center of the
parabolic trap. However, for $E > 4J$ particles are Bragg-scattered
by the lattice, and perform Bloch-like-oscillations centered far from the trap
center~\cite{RigolPRA}. The particle localization corresponds to the population
of these latter high-energy modes, which becomes significant for displacements
$\delta \gtrsim \delta_{\rm c}$,~\cite{AnaPRA05}.
Our numerical results in the limit of weak interactions are shown
in Fig.~\ref{fig:fig1}(a-c), where dipole oscillations of the center of mass  $x_{\rm com}$ are
shown as a function of time $t$, for different values of the displacement $\delta$.
In the simulations, as initial condition we use the ground-state wavefunction
of the undisplaced potential, shifted by $\delta$ lattice sites. 
On increasing $\delta$, the dynamics changes from undamped
to damped, and the particles oscillate around the trap center. On increasing further the displacement
[$\delta\gtrsim 5$ in panels (a-b)] the oscillations are overdamped, and the COM slowly drifts towards
the trap center or clings to the borders of the trap [case with $N=23$ of panel $c)$].
This behavior corresponds to the localization transition predicted by mean-field theory.
However, Fig.~\ref{fig:fig1} shows that quantum fluctuations, properly
accounted for by the tDMRG, smear out the transition into a smooth crossover
between the undamped and the overdamped regimes. 

Having established a connection with known results in the mean-field regime,
we now present exact results for the particle localization in the interesting
case of stronger interactions $U/J \gtrsim 1$ and $\delta \lesssim \delta_{\rm c}$. 
We first focus on model systems and fix $\delta_{\rm c}=6$ and the
displacement $\delta=1 < \delta_{\rm c}$, such that for small interactions
$U/J \lesssim 1$ the dynamical instability discussed above does not occur,
e.g. for $U/J=1$ the dipole oscillations are undamped for all $N$, see
Figs.~\ref{fig:fig1}(a)-(b). The dipole dynamics is then studied as a function
of the ratio $U/J$. 
In particular, Fig.~\ref{fig:fig2}(a) shows the damping rate $\Gamma$ of the
dipole oscillations as a function of $U/J$ for $N=11, 15$ and 28 [exemplifying
  cases $a),b)$ and $c)$ above]. 
Here, $\Gamma$ is calculated using the expression for underdamped oscillations
$x_{\rm com}(t)= e^{-\Gamma t}[1-\cos(\Omega t + \phi_0)]+y_0$,
with $\Gamma$, $\phi_0$ and $y_0$ fitting parameters.
Three key observations are in order.
i) The damping rate {\em increases exponentially} with $U/J$ for intermediate
interaction strengths $2 \lesssim U/J \lesssim 6$, a result which is not
captured by mean-field, and  is significantly larger than what predicted using
phase-slip techniques, valid for $U\lesssim 1$~\cite{Polko,AnaPRA05}.
ii) Eventually for large enough interactions ($U/J \sim 6$) the oscillations are overdamped for all $N$.
We find that for the cases $N=15$ and $28$, this overdamping corresponds
to the {\em formation of a Mott-state} and a {\em cake-structure} as in
Fig.~\ref{fig:fig0}(b) and (c), respectively. In particular, for $N=15$ the
particle localization occurs for $U/J \approx 4$, a value remarkably close to
the superfluid/Mott-insulator quantum phase transition in an homogeneous
lattice at commensurate filling and zero current. That is, the results for
$\delta < \delta_{\rm c}$ naturally interpolate between the finite-current
dynamical instability and the zero-current quantum phase
transition~\cite{SmerziPRL}.
iii) Despite the Mott-formation for large $N$, for a given $U/J$ the damping
$\Gamma$ is actually larger for {\em smaller N}, such that for $N=11$ the
dynamics is frozen already for $U/J < 4$. In the following we show that this
has crucial consequences for the interpretation of the results of
Ref.~\cite{fertig05} in the most interesting regime of interactions $U/J\sim
4$. 

In the experiment of Ref.~\cite{fertig05}, the decay of dipole oscillations was studied as a function of the
optical lattice depth $V_0$ for a fixed displacement $\delta=8$, finding
damping already for weak lattices $V_0/E_R > 0.5$, with $E_R$ the recoil
energy. The experimental data are shown as black dots in
Fig.~\ref{fig:fig2}(b) as a function of $V_0$ in the range $2 \lesssim V_0/E_R
\lesssim 5$, where the use of Eq.~\eqref{HamBHInt} is
justified~\cite{Jaksch98,AnaPRA05}, corresponding to the  interesting regime
of interactions $3 \lesssim U/J \lesssim 8$. For $V_0/E_R = 3$ and $V_0/E_R >
3$ the value of the damping rate $\Gamma$ has been extracted using formulas
appropriate for underdamped and overdamped motion,
respectively~\cite{fertig05}. 
The most interesting experimental finding shown in Fig.~\ref{fig:fig2}(b) is
the measurement of an abrupt transition from a weakly damped regime to an
overdamped regime for a lattice depth $V_0/E_R \simeq 3$, where the damping
rate $\Gamma$ of the dipole oscillations increases by more than an order of
magnitude. The physical mechanism behind this apparent transition has proven
elusive. 

In Fig.~\ref{fig:fig2}(b) the experimental results are compared to our
numerical results for $N=80$ and 45, green diamonds and red squares,
respectively. The value $N=80$ has been chosen since it corresponds to the
number of particles in the central 1D lattice of the array in the experiment,
which is the most largely populated with $\langle n_j \rangle > 1$ for all
$U/J$, as in Fig.~\ref{fig:fig0}(c). Conversely, the case $N=45$ exemplifies
case (b), with $\langle n_j \rangle \lesssim 1$ for $U/J \gtrsim 4$. The
figure shows a very good agreement between the numerical and the experimental
results in the entire region $2\lesssim V_0/E_R \lesssim 5$ ($3 \lesssim U/J
\lesssim 8$). However, 
the case $N=80$ slightly underestimates the damping around $V_0/E_R \simeq 4$, while the agreement for
$N=45$ is almost perfect. For $V_0/E_R \gtrsim 5$ all numerical results fall
inside the experimental errorbars, however, the case $N=45$ shows a strong
damping, while the case $N=80$ falls in the middle of the experimental
errorbars. The explanation of the results above stems from the observation
that in the experiment $\delta_{\rm c}$ varies between $\delta_{\rm c} \sim
18$ and 15 for $3 \lesssim V_0/E_R \lesssim 5$, and thus $\delta < \delta_{\rm
  c}$ for all lattice depths. We can then use the results for the model
systems of Fig.~\ref{fig:fig2}(a) to explain the experimental findings. That
is: i) the transition observed experimentally at $V_0/E_R\simeq 3$ is actually
a {\em crossover}, where the 1D systems with the lowest number of particles
tend to localize first, in agreement with the discussion of
Fig.~\ref{fig:fig2}(a). ii) For $V_0/E_R \gtrsim 5$, the dynamics of particles
in the 1D systems with $\langle n_j \rangle\leq 1$ ($N=45$ in the simulations)
is completely frozen, and the overall mobility of the cloud is due to residual
oscillations 
in lattices with higher onsite density. This latter observation is in
agreement with the results of Ref.~\cite{PupilloNJP06}, where it is shown that
for $V_0/E_R>5$ the damping rate observed in the experiment is well reproduced
by the results for $N=80$. We notice that numerical results for $N=80$
consistent with ours have been recently reported in~\cite{Clark08}, however
the focus here is on a comprehensive explanation of the
experiment~\cite{fertig05}. 

The different behaviors of $\Gamma$ for $N=45$ and 80 and $U/J > 4$ can be
modeled as follow. In the low-density case with $N=45$ the tendency to
localization is explained by noting that interactions broaden the spatial
width of the atom cloud, until  
the onsite density falls below one [see also Fig.~\ref{fig:fig0}(a-b)]. In
this case, the low-energy physics maps into that of an extended cloud of
non-interacting fermions, with single-band Hamiltonian~\cite{AnaPRA05} 
\begin{eqnarray}
\tilde H_1(t) = - J \sum_{<i,j>} c^{\dagger}_i c_j +  \Omega \sum_j [j-\delta(t)]^2 c^{\dagger}_j c_j, \nonumber
\end{eqnarray}
with $c_j$ and $c^{\dagger}_j$ fermionic operators. For large enough
displacements $\delta$, the fermions largely occupy localized modes of the
single-particle spectrum discussed above, and the COM remains frozen. 
The dynamics of interacting particles at large density, e.g. $N=80$ in
Fig.~\ref{fig:fig2}(b), can be modeled starting from the case of largest
interactions $U/J\gg 1$, where the density profile has a cake-like structure,
Fig.~\ref{fig:fig0}(c). 
This situation is well described by an extended fermionization model~\cite{PupilloNJP06,PupilloPRA06,Popp06}, where
Eq.~\eqref{HamBHInt} is replaced by an effective Hamiltonian with two coupled Fermi bands separated
by an energy $U$~\cite{Popp06}
\begin{eqnarray}
\tilde H_2(t) = - J \sum_{<i,j>} [c^{\dagger}_i c_j + 2 d^{\dagger}_i d_j + \sqrt{2}  (c^{\dagger}_i d_j + d^{\dagger}_i c_j )]\nonumber\\
 + \sum_j (\Omega [j+\delta(t)]^2 c^{\dagger}_j c_j + \{\Omega [j+\delta(t)]^2 + U\} d^{\dagger}_j d_j),\label{Eq:EqHamEff}
\end{eqnarray}
with the operators $c_j$, $c^{\dagger}_j$ and $d_j$, $d^{\dagger}_j$ referring to the lower and
higher energy bands of width $4J$ and $8J$, respectively. Oscillations in this limit
are due to the dynamics of the (delocalized)
$d_j$-fermions of Eq.~\eqref{Eq:EqHamEff} in the higher-energy band, while $c_j$-fermions are frozen in a (band) insulator. Observing
these residual oscillations thus corresponds to probing the superfluidity
of bosons with two-particles per site in a homogeneous lattice, in a local-density-approximation sense~\cite{Fisher}.
This picture, valid for $U/J \gg 1$~\cite{PupilloNJP06,Popp06},
can be extended to gain a qualitative insight in the dependence of the dipole oscillations on interactions for
$4 \lesssim U/J \lesssim 10$. In fact, neglecting the parabolic potential, in this regime the
model of Eq.~\eqref{Eq:EqHamEff} suggests that the spectrum is continuum, since the gap $U$ between the
two Fermi bands is smaller than their total width. It is thus plausible that Bloch-like
oscillations of the particles are here suppressed, and transport restored.
However, for $U \gtrsim 12 J$ the energy spectrum develops a gap again around $4J$, and thus transport
in the lower-energy band is inhibited. Residual current is then due to delocalized particles in
the higher-energy band, as explained above. We notice that this picture is consistent with our numerical
findings for $U/J > 5$ in Fig.~\ref{fig:fig2}(b).

In conclusion, we have explained the experiment in~\cite{fertig05} in the most
interesting regime of intermediate interactions. The very good agreement
between experimental and tDMRG results demonstrates the latter as a unique
tool for quantitative comparisons with cold gases experiments in the strongly
correlated regime in one dimension. 

Discussions with A.M. Rey, C.J. Williams and C.W. Clark are gratefully
acknowledged. This work was supported by OLAQUI, NAMEQUAM, FWF, MURI, EUROSQIP
and DARPA and developed using the DMRG code released
within the PwP project (www.dmrg.it).

\end{document}